# Estimation of Reconnection Flux Using Post-eruption Arcades and Its Relevance to 1-AU Magnetic Clouds

N. GOPALSWAMY

*NASA Goddard Space Flight Center, Greenbelt, MD, USA*

S. YASHIRO, S. AKIYAMA, H. XIE

*The Catholic University of America, Washington, DC, USA*

Abstract

We report on a new method to compute the flare reconnection (RC) flux from post-eruption arcades (PEAs) and the underlying photospheric magnetic fields. In previous works, the RC flux has been computed using cumulative flare ribbon area. Here we obtain the RC flux as half of that underlying the PEA associated with the eruption using an image in EUV taken after the flare maximum. We apply this method to a set of 21 eruptions that originated near the solar disk center in Solar Cycle 23. We find that the RC flux from the arcade method ($\Phi_{rA}$) has excellent agreement with that from the flare-ribbon method ($\Phi_{rR}$) according to: $\Phi_{rA} = 1.24(\Phi_{rR})^{0.99}$. We also find $\Phi_{rA}$ to be correlated with the poloidal flux ($\Phi_P$) of the associated magnetic cloud at 1 AU: $\Phi_P = 1.20(\Phi_{rA})^{0.85}$. This relation is nearly identical to that obtained by Qiu *et al.* (*Astrophys. J.* **659**, 758, 2007) using a set of only nine eruptions. Our result supports the idea that flare reconnection results in the formation of the flux rope and PEA as a common process.

## 1. Introduction

A number of investigations have identified a close connection between coronal mass ejections (CMEs) and the associated flares: (i) the CME acceleration is synchronized with the rise time of the associated flare (Zhang *et al.*, 2001; Zhang and Dere, 2006; Gopalswamy *et al.*, 2012), (ii) the CME kinetic energy and soft X-ray peak flux are correlated (Gopalswamy, 2009), (iii) the CME width is determined by the flare magnetic field (Moore, Sterling, and Suess 2007), (iv) flare reconnection (RC) and flux rope formation are related (Leamon *et al.*, 2004; Longcope and Beveridge, 2007; Qiu *et al.*, 2007; Hu *et al.*, 2014), (v) the CME nose is directly above the flare location (Yashiro *et al.*, 2008), and (vi) the high charge state of minor ions in interplanetary coronal mass ejections (ICMEs) is a consequence of the heated flare plasma entering into the CME flux rope during the eruption (Lepri *et al.,* 2001; Reinard, 2008; Gopalswamy *et al.,,* 2013). One of the key aspects of CMEs is their flux rope nature considered extensively from theory and observations (see *e.g.*, Mouschovias and Poland 1978; Burlaga *et al.,* 1981; Marubashi 1997; Gibson *et al.,,* 2006; Linton and Moldwin 2009). The flux rope nature of CMEs provides an important eruption scenario that can be tested using remote-sensing and *in-situ* observations.

A number of investigations have also shown that the flare RC process results in the simultaneous formation of a post-eruption arcade (PEA) and a flux rope during solar eruptive events (Leamon *et al.,* 2004; Longcope and Beveridge, 2007; Qiu *et al.,* 2007; Hu *et al.,* 2014). At present it is not fully understood whether CME flux ropes exist before eruption or formed during eruption. The reality may be something in between: flux may be added to a pre-existing flux rope via flare RC (see *e.g.*, Lin, Raymond, and van Ballegooijen, 2004). If the flux rope is formed due to reconnection, for each flare loop formed, the flux rope gets one turn, as explained in Longcope *et al.,* 2007; Longcope and Beveridge (2007). The flare loops are rooted in the flare ribbons on either side of the polarity inversion line. Thus flare reconnection flux $\Phi_r$ and the poloidal flux $\Phi_p$ of the associated flux rope at 1-AU are expected to be equal. On the other hand $\Phi_r < \Phi_p$ might indicate that flux is added to preexisting flux of the flux rope. For a set of 9 eruptions, Qiu *et al.* (2007) computed $\Phi_r$ from the *Transition Region and Coronal Explorer* (TRACE: Strong *et al.*, 1994) 1600 Å flare-ribbon areas and photospheric/chromospheric magnetic field strength. They compared $\Phi_r$ with $\Phi_p$ obtained by fitting a flux rope to the *in-situ* observations of the associated CMEs and found an approximate equality between $\Phi_r$ and $\Phi_p$. Qiu *et al.* (2007) analysis resulted in the relation,

$$\Phi_p = 1.12(\Phi_r)^{0.82}, \quad \ldots\ldots\ldots\ldots\ldots\ldots\ldots\ldots\ldots\ldots\ldots\ldots\ldots\ldots \quad (1)$$

suggesting that $\Phi_r$ is approximately equal to $\Phi_p$. One of the reasons for a small number of events is the scarce observations of flare ribbons. The purpose of this paper is to report on a new technique to compute $\Phi_r$ from PEAs and also to test the $\Phi_r - \Phi_p$ relationship with a larger data set.

## 2. Description of the New Technique and Data

The post-eruption arcade (PEA) technique to determine $\Phi_r$ makes use of the close relationship between flare ribbons and PEAs because the ribbons essentially mark the footprints of PEAs. Instead of counting the ribbon pixels in a series of images, we simply demarcate the area under PEAs, $A_P$, using a polygon, overlay the polygon on a magnetogram obtained around the time of the PEA image, and sum the absolute value of the photospheric magnetic flux in all the pixels within the polygon. The resulting RC flux is then half of the total flux through the polygon. The justification for this technique is that the ribbon separation ends after the flare maximum and the accumulated pixel area should roughly correspond to the area under the PEA on one side of the polarity inversion line.

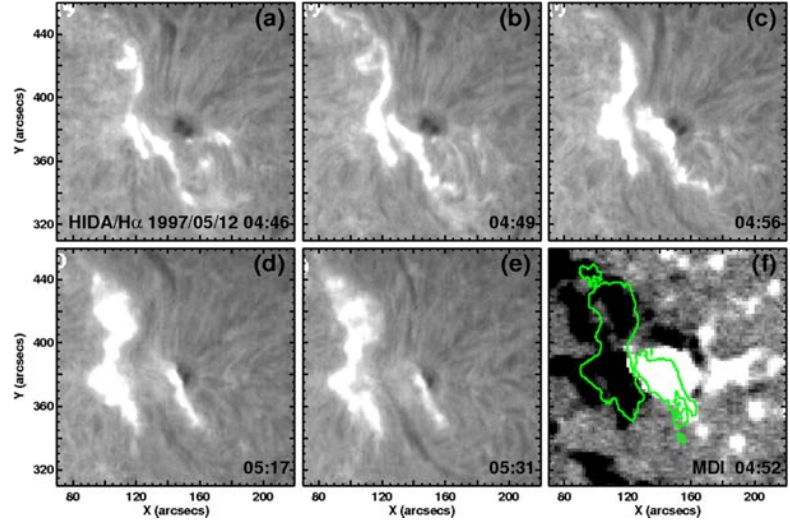

Figure 1. A series of Hα images (a-e) from the Hida Observatory of Kyoto University in Japan showing flare ribbons for the 12 May 1997 (event No. 2) eruption from 04:45 to 05:31 UT. The cumulative ribbon areas are denoted by the green contours in panel (f) is superposed on an MDI magnetogram obtained at 04:52 UT (white is positive and black is negative magnetic field in the magnetogram). The total ribbon area is $1.7 \times 10^{19}$ cm$^2$. With the average field strength of 171.5 G within the ribbon area, the RC flux becomes $1.46 \times 10^{21}$ Mx (average under one ribbon).

## 2.1 Illustrative Examples

As an example we consider the 12 May 1997 eruption, which resulted in a magnetic cloud (MC) detected by the *Wind* and *Advance Composition Explorer* (ACE) spacecraft (Gopalswamy *et al.*, 2010). The eruption occurred in NOAA active region (AR) 8038 located at N21W08. Various aspects of this event have been studied extensively using remote observations (Thompson *et al.*, 1998), *in-situ* observations (Baker *et al.*, 1998), and modeling (Titov *et al.*, 2008). Figure 1 shows the evolution of the eruption in a series of Hα pictures from the *Domeless Solar Tower Telescope* of Kyoto University at Hida Observatory in Japan (Nakai and Hattori, 1985) revealing the evolution of the flare ribbons. Initially the ribbons are thin and close to each other. As time progresses, the ribbons separate and their areas increase. The RC flux in the ribbons is the magnetic flux computed over the cumulative area of the ribbons on one side of the polarity inversion line (PIL) in the flaring region. Ideally, the ribbons are well defined over the whole event and one can use either side of the PIL. In practice, the ribbons can have different morphology on either side of the PIL. If the measurements are from both sides of the ribbon, we divided the flux by two. Sometimes, we were able to measure the area only from one side of the PIL. In computing the cumulative area, we eliminated the overlapping area in successive frames to avoid double counting. In the case of the May 12 1997 event, both ribbons are well defined, so we compute the total cumulative area under both ribbons. The outline of the cumulative ribbon areas are superposed over a magnetogram obtained by the *Michelson Doppler Imager* (MDI: Scherrer *et al.,* 1995) on board the *Solar and Heliospheric Observatory* (SOHO: Domingo *et al.,* 1995) as shown in Figure 1. The cumulative ribbon area was $1.7\times10^{19}$ cm$^2$ (both ribbons) and the average field strength was 171.5 G. The RC flux corresponds to the flux in one ribbon, so we get an average $\Phi_{rR} = 1.46\times10^{21}$ Mx. We denote the RC flux from the ribbon method by $\Phi_{rR}$ to distinguish it from the one obtained using the arcade method ($\Phi_{rA}$).

In order to compute $\Phi_{rA}$, we consider a 195 Å EUV image at 16:50 UT obtained by the *Extreme-ultraviolet Imaging Telescope* (EIT: Delaboudiniére *et al.,* 1995) on board SOHO. By visual examination, we marked the edges of the PEA as a polygon, which we superposed on an MDI magnetogram obtained at 16:04 UT (see Figure 2). The flux in each MDI pixel within the polygon is summed to get the total flux in the arcade area as $4.72\times10^{21}$ Mx. Half of this quantity is the flare RC flux from the arcade method, $\Phi_{rA} = 2.36\times10^{21}$ Mx. We see that $\Phi_{rA}$ is larger than $\Phi_{rR}$ by 38%. Given the uncertainties in identifying the edges of the ribbons, the agreement is reasonable.

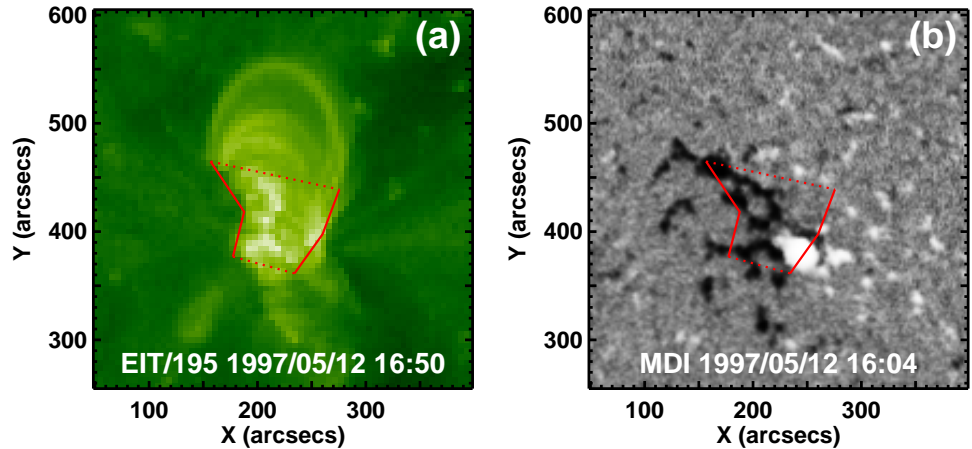

Figure 2. (a) The post-eruption arcade imaged by SOHO/EIT at 195 Å at 16:50 UT for the 12 May 1997 (event No. 2) eruption. The solid lines are drawn near the feet of the arcade, while the dashed lines complete the polygon. The EUV images have a pixel size of ~2.6 arcsec. (b) The polygon from panel (a) is superposed on an MDI magnetogram at 16:04 UT. Summing the flux from each pixel within the area occupied by the arcade (corrected for projection effects) and diving by 2, we get the total flux as $2.36 \times10^{21}$ Mx (the arcade area is $4.25\times10^{19}$ cm$^2$).

Since there were not many events with good Hα flare observations, we also considered flare ribbons observed with TRACE at 1600 Å. Figure 3 shows an event with PEAs from SOHO/EIT 195 Å and TRACE 171 Å, while the ribbons are from TRACE 1600 Å. The eruption is the famous Bastille Day (14 July 2000) event, which had severe space weather impact because of a large solar energetic particle event including ground level enhancement (Gopalswamy *et al.,* 2004; Gopalswamy *et al.,* 2012; Mewaldt *et al.,* 2012) and a super-intense geomagnetic storm (Zhang *et al.,* 2007). Various aspects of this eruption have been extensively studied (Reiner *et al.,* 2001; Aschwanden and Alexander, 2001;Yan and Huang 2003 and references therein). As in Figure 1, we mark the PEA area by a polygon drawn on the SOHO/EIT 195 Å image. The TRACE image has a spatial resolution of ~1 arcsec compared to ~5 arcsec in the EIT image, which is evident in the detailed structure of the arcade in the TRACE image. We also see some difference at the edges of the arcade where TRACE observes additional faint structures. Despite this difference, we use the EIT images for identifying the arcade area because they are full disk images and hence most of the eruptions are observed. Furthermore, not all arcades were observed by TRACE because of its limited FOV and was generally pointed at intense flaring regions. The RC flux estimated from the EIT area of the arcade and the MDI pixels within the area gives a flux of $13.1 \times 10^{21}$ Mx. The additional structures to the right edge of the polygon contribute only $\sim 5.2 \times 10^{19}$ Mx, which is only about 0.4% of the arcade flux and hence negligible. Figure 3 also shows that the ribbon is very extensive and well defined for measurements of the RC flux from flare ribbons. We used 71 of the 381 TRACE 1600 Å images available between 10:00 UT and 14:59 UT to obtain the ribbon area as $5.64 \times 10^{19}$ cm$^2$. The average magnetic field strength under the ribbon area was 446.9 G. The average RC flux from the ribbons on one side of the neutral line was obtained as $12.61 \times 10^{21}$ Mx. This value is also close to the RC flux from the arcade method, differing only by ~3.7%. Note that the RC flux in the 14 July 2000 event is much larger than that in the 12 May 1997 event because of the high magnetic field strength. Only one Hα image at 10:03 UT was available for this event. The ribbon area was $1.69 \times 10^{19}$ cm$^2$ and the average field strength was 557.5 G, giving a partial RC flux of $9.44 \times 10^{21}$ Mx, which is an underestimate, but consistent with the RC flux from the arcade method.

## 2.2 The Data Set

We are interested in solar eruptions that result in a MC at Earth, so we can make quantitative comparison between the flare RC flux and the poloidal flux of the associated MCs. For this purpose, we started with the list of 54 eruptions in Solar Cycle 23 considered for the Flux Rope CDAW workshops (Gopalswamy *et al.,* 2013). The eruptions occurred from within ±15° in longitude from the disk center. The longitudinal criterion was imposed to make sure the associated CMEs observed by the *Large Angle and Spectrometric Coronagraph* (LASCO: Brueckner *et al.,* 1995) arrived at Earth as interplanetary CMEs (ICMEs) as detected *in situ* by spacecraft located at Sun-Earth L1. The PEAs in the eruptions were observed in EUV or X-rays and their measureable properties have already been reported (*e.g.*, Yashiro *et al.,* 2013). These authors compared PEAs associated with CMEs that ended up at 1 AU as MCs and non-cloud ICMEs. We are interested in MCs because their flux rope structure allows us to determine the poloidal flux of the MCs at 1 AU (*e.g.*, Lepping *et al.,* 1990) to compare them with the RC flux. The fitted parameters of the MCs are available online (http://wind.nasa.gov/mfi/mag_cloud_S1.html).

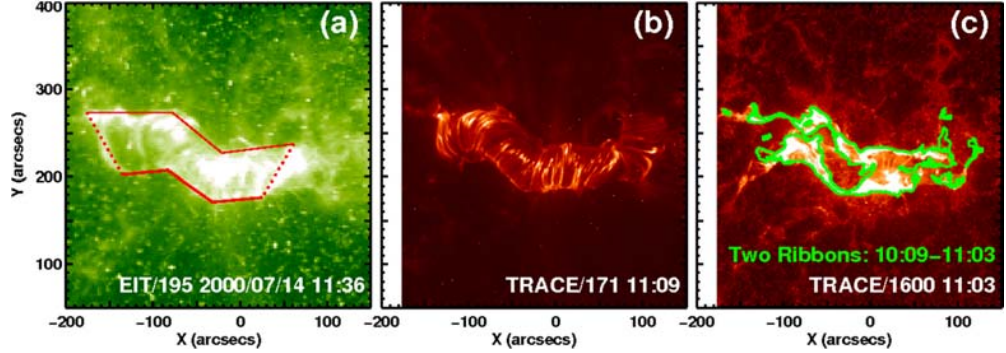

Figure 3. PEA of the 14 July 2000 (event No. 19) eruption as observed by (a) SOHO/EIT at 195 Å and (b) TRACE at 171 Å. The TRACE image in panel (c) taken at 10:39 UT shows the ribbons at 1600 Å. The green contour in (c) corresponds to the cumulative ribbon area (TRACE 1600 Å) between 10:00 and 14:59 UT.

Of the 54 ICMEs, only 23 were MCs. For one of the MCs, there was no EIT observations. For another event, the arcade at the Sun was not well defined. Excluding these two events, we have listed the remaining 21 events in Table 1. The event No. (column 1) is from the CDAW list (Gopalswamy *et al.,* 2013). We have retained the original event identifiers for easy comparison. The date and time of the MC and the associated CME at the Sun are listed in columns 2 and 3, respectively. The start time of the associated soft X-ray flare, the flare class, and the flare location in heliographic coordinates are listed in columns 4, 5, and 6, respectively. The area under the EUV arcade ($A_a$ in units of $10^{19}$ cm$^2$, column 7), the average field strength ($B_a$ in Gauss, in column 8) and half of the total unsigned magnetic flux under the arcade ($\Phi_{rA}$ in units of $10^{21}$ Mx, column 9) are from the PEA method described in Figures 1 and 2. Hα observations were generally not uniform and were also not available for many events. In some cases, we were able to measure the ribbon area only on one side of the neutral line. In some cases, only a single frame of the Hα image was available; the ribbon area in such cases represent a lower limit to the area. In all, there was at least one Hα picture for 12 events. Of these, only eight had more than one Hα frames; the number of frames were sufficient to obtain the RC flux only in six cases. For the remaining events, the ribbon areas and hence the RC flux were underestimated. The ribbon area ($A_R$, in units of $10^{19}$ cm$^2$), the average magnetic field strength ($B_R$ in units of Gauss), and the RC flux from the ribbon method are listed in columns 10, 11, and 12, respectively. When we searched for ribbon observations in TRACE 1600 Å data, we found eight events (Nos. 19, 21, 32, 43, 45, 46, 49, and 53 noted with a superscript "T" in column 10). For event No. 43 there was only one frame in the rise phase, so the ribbon area is an underestimate. The remaining seven events had usable TRACE data. In column 10, the source of ribbon data (H – Hα; T – TRACE) is noted. The poloidal flux $\Phi_p$ of MCs computed from the Lundquist solution (Lepping *et al.,* 1990) is listed in column 13. Of the many output parameters obtained by the flux rope fitting, we use the axial field strength ($B_0$) and the flux rope radius ($R_0$) at 1-AU to get the poloidal field of the MC,

$$\Phi_p = L\,(B_0 R_0)/x_{01}, \dots\dots\dots\dots\dots\dots\dots\dots\dots\dots\dots\dots\dots\dots\dots\dots\dots\dots\dots\dots\dots\dots\dots\dots..(2)$$

supposing that the flux rope extends up to the radius where the axial field component vanishes. Here $x_{01}$ is the first zero (2.4048) of the Bessel function $J_0$ and $L$ is the total length of the flux rope, taken as 2 AU following Nindos, Zhang, and Zhang (2003).

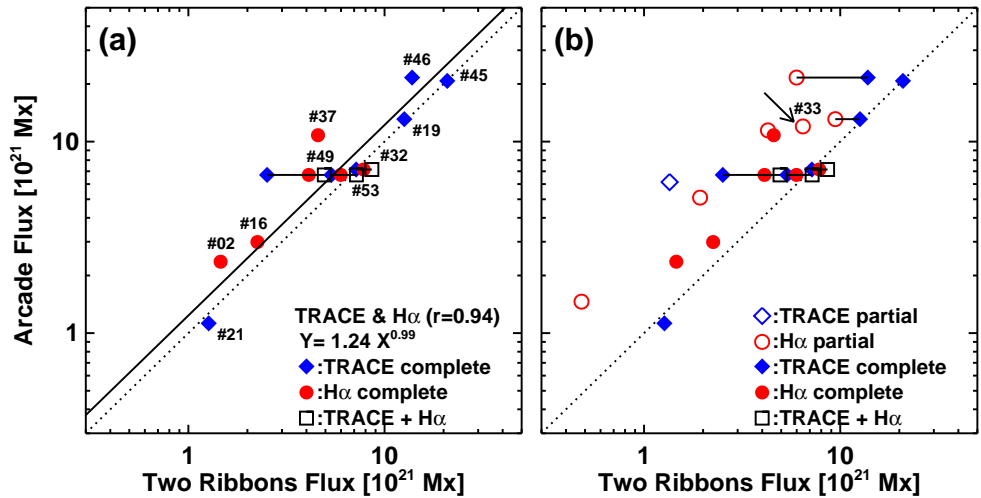

Figure 4. (a) Scatter plot between the RC fluxes obtained from the arcade method and the ribbon method for ten events identified by the event numbers. The RC flux from Hα ribbons (red symbols) and TRACE 1600 Å ribbons (blue symbols) are distinguished. Data points connected by horizontal line represent ribbon measurements from Hα and TRACE. In these cases, the TRACE and Hα data were combined in getting the cumulative area and the resulting flux is indicated by a black square (event Nos. 32, 49, and 53). The dotted line corresponds to equal arcade and ribbon fluxes. The solid line is the least squares fit to the data points (Hα + TRACE 1600 Å). The correlation coefficient (0.94) is higher than the critical value (0.549) of the Pearson correlation coefficient for ten data points at 95% confidence level. In events with both Hα and TRACE measurements, combined flux (black squares) is used in the correlation. (b) The same scatter plot, but includes events with incomplete ribbon data (open symbols). The plotted fluxes are lower limits for these events. Event No. 33 discussed in the text is pointed by an arrow.

## 3. Analysis and Results
### 3.1 Comparison between RC flux from the Arcade and Ribbon Methods

We extend the case studies presented in Section 2.1 to all cycle-23 eruptions that originated within ±15º from the disk center and resulted in MCs at 1 AU (see Table 1). First we check if the arcade and ribbon methods yield consistent results. To do this we have extracted the 15 events with ribbon information (Hα or TRACE) and listed them in Table 2. For six events, there were no usable ribbon data either from Hα or from TRACE 1600 Å (Nos. 23, 24, 27, 39, 44, and 54 in Table 1). We have separated Hα and TRACE data on ribbons along with the number of frames available for each event. We have also listed the number of frames available for each event in Hα and TRACE. In the last column, we have listed the name of the observatory that provided the Hα images. If the number of frames is insufficient to make complete ribbon measurements in an event, we considered the computed RC flux be an underestimate. Only three events (Nos. 32, 49 and 53) had data both in Hα and TRACE. Table 2 shows that, there were only a total of ten events with RC flux computed from flare ribbons. In the remaining five events, ribbons were observed, but not in sufficiently large number of frames to compute the cumulative ribbon areas. However, we use such events to show that they provide lower limits to the RC flux and check if they have the correct trend.

The RC fluxes from the arcade and ribbon methods are shown in Figure 4a as a scatter plot, indicating a high correlation (r = 0.94). A least squares fit gives the regression equation,
$$\Phi_{rA} = 1.24(\Phi_{rR})^{0.99} \ldots\ldots\ldots\ldots\ldots\ldots\ldots\ldots\ldots\ldots\ldots\ldots\ldots\ldots\ldots\ldots\ldots\ldots\ldots\ldots\ldots\ldots\ldots(3)$$
This result is significant because it shows that the simpler arcade method works well. The regression line deviates only slightly from the equal-fluxes line ($\Phi_{rA} = \Phi_{rR}$). The largest deviation is for the event No. 37, in which $\Phi_{rA}$ is greater than $\Phi_{rR}$ by a factor of ~2.4. In Figure 4b, we have included events that did not have complete ribbon information (open symbols). The RC fluxes from the ribbon method are underestimated in these events due to insufficient number of frames available. However, the open symbols are consistent with the trend that the RC flux from the arcade method correlates with that from the ribbon method. If these events had complete observations, the open symbols would move closer to the equal fluxes line. We also note that five of the ten data points in Figure 4a are on the line of equal fluxes. The fact that the equal fluxes line needs to be multiplied by 1.24 to get the regression line indicates an overestimate of the RC flux by the arcade method, an underestimate by the ribbon method or both. The possibilities for the overestimate of the RC flux from the arcade method include: (i) the arcade is observed in the corona, resulting in a slight overestimate of the arcade area compared to the actual area at the photospheric level and hence a higher RC flux, and (ii) the arcade flux might include contributions from very close to the neutral line, while the ribbons may start from a finite distance from the neutral line. Such fluxes do not participate in the flare reconnection can also contribute to an overestimate of the arcade flux.

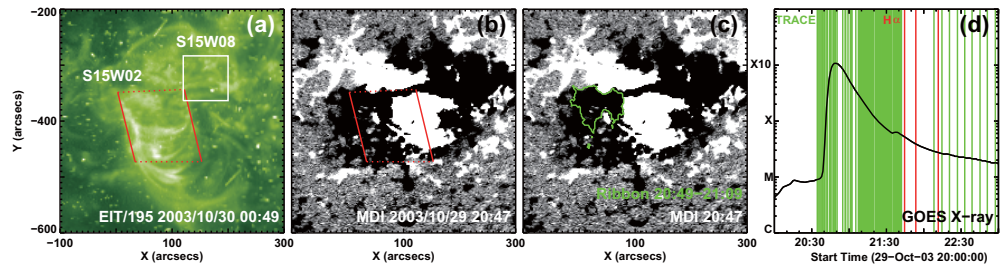

Figure 5. The 29 October 2003 eruption (No. 46) that showed a large deviation between RC fluxes from the arcade and ribbon methods for the group of ten cases in which $\Phi_{rR}$ could be estimated. (a) PEA at 00:49 UT on 30 October 2003 with the edges of the arcade marked by the red lines. The white box placed at S15W08 marks the area where another flare occurred, but it did not affect the flux computations. (b) The arcade area superposed on an MDI magnetogram taken at 20:47 UT. (c) The combined ribbon area superposed on the MDI magnetogram. (d) The GOES soft X-ray light curve (black) with the times of TRACE (green) and Hα (red) images used in computing the ribbon areas. The second flare can be seen as a small increase in the GOES intensity around 21:30 UT. The main part of the flare had high-cadence TRACE observations, while in the late decay phase the cadence dropped to below 10 min. There were only three Hα images available for this event.

Event No. 37 had 11 Hα frames covering most of the flare duration. There was no data coverage for the last 2 hours in the decay phase of the flare. Most likely, the RC flux from the ribbon method was underestimated. Except for this event, we see from Table 2 that the ratio $\Phi_{rA}/\Phi_{rR}$ is in the range 0.89 to 1.6 for all events having $\Phi_{rR}$ well defined. One of the events with the largest ratio (~1.6) is the 29 October 2003 eruption (No. 46). This well-known eruption is from AR 10486 that resulted in a magnetic cloud (Gopalswamy *et al.*, 2005). Figure 5 shows an overview of the event with the EUV arcade and the PIL involved. The ribbon on the positive side was fragmented because of a narrow lane of negative field region at the northern end of the arcade. We used the ribbon on the negative side of the PIL, which was not fragmented. The GOES soft X-ray plot in Figure 5 has the times of the available Hα and TRACE 1600 Å frames marked; these frames were used to compute the RC flux using the ribbon method. The early part of the flare ribbons were observed with high cadence, while the late decay phase was observed with lower cadence by TRACE. In Hα, there were only three frames in the decay phase. Using the TRACE observations alone, we obtained a RC flux of ~ $13.82 \times 10^{21}$ Mx. The Hα observation alone gave a smaller value ($5.99 \times 10^{21}$ Mx) because of the incomplete observations. On the other hand, the $\Phi_{rA}$ of this event was $21.6 \times 10^{21}$ Mx. It seems unlikely that $\Phi_{rA}$ in this event is overestimated. An underestimate of the RC flux using the ribbon method is possible because of the uneven coverage. For example, there was a gap of ~ 30 min between the last TRACE observation and the first Hα observation. Figure 5 also shows that the arcade extends further to the south than the cumulative ribbon area does, suggesting possible underestimate of the RC flux. Thus, using the ribbon method can be subjective depending on the threshold used in defining the ribbons. The other Halloween event on 28 October 2003 (No. 45) overlapped with Qiu *et al.* (2007) list, who estimated the RC flux from TRACE ribbons as $18.8\times10^{21}$ Mx, which is similar to our value ($20.88\times10^{21}$ Mx). Both are in good agreement with the RC flux from the arcade method.

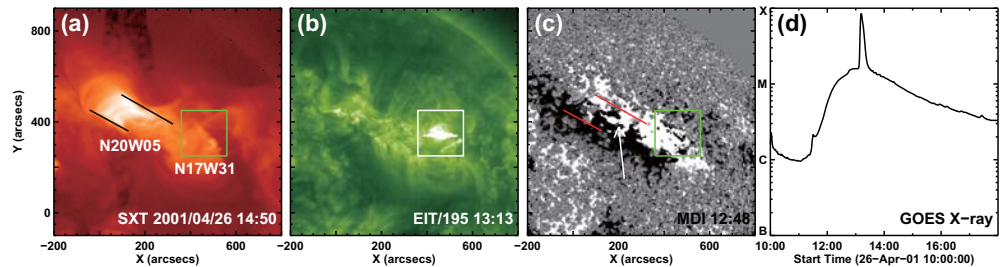

Figure 6. (a) A *Yohkoh/Soft X-ray Telescope* (SXT) image showing the PEA associated with the 26 April 2001 (No. 33) eruption. The arcade was associated with the M1.5 gradual flare from N20W05. There was another impulsive flare starting around 13:04 UT, but to the west at N17W31 (inside the box shown). The full disk SXT images did not capture the impulsive event, but the SOHO/EIT image shows the compact event at 13:13 UT enclosed by the box in (b). (c) MDI magnetogram of the extended magnetic region that hosted both the gradual and the impulsive eruptions. NOAA AR 9433 (pointed by arrow) was located between the two eruptions, but did not participate in the eruption. The feet of the arcade (red lines) and the box within which the impulsive flare occurred are superposed on the magnetogram. (d) The GOES soft X-ray light curve showing the gradual M1.5 flare and the impulsive M8.9 flare.

A clear case of overestimation of the arcade flux was in the 26 April 2001 (No. 33) event, which is marked in Figure 4b. A large bipolar active region was located near the PIL of the arcade, but did not participate in the flare reconnection and hence contributed to the overestimate of $\Phi_{rA}$. We estimate the unsigned flux due to the bipole as $\sim15.2\times10^{21}$ Mx from the MDI magnetogram. Since the bipole did not participate in the eruption process, it needs to be subtracted from the total arcade flux ($39.2\times10^{21}$ Mx). The corrected arcade flux is $24.0\times10^{21}$ Mx, so the RC flux is $12.0\times10^{21}$ Mx. The corrected value is now closer to the equal fluxes line. Unfortunately there were Hα frames only in the decay phase of the flare, so the estimated RC flux is an underestimate. Using the nine frames available, the RC flux from the ribbon method was estimated as $6.45\times10^{21}$ Mx. The true value is expected to be higher, so the data point will move to the right, closer to the equal fluxes line. This event demonstrates that large deviations can be understood by examining the magnetograms and the PEAs.

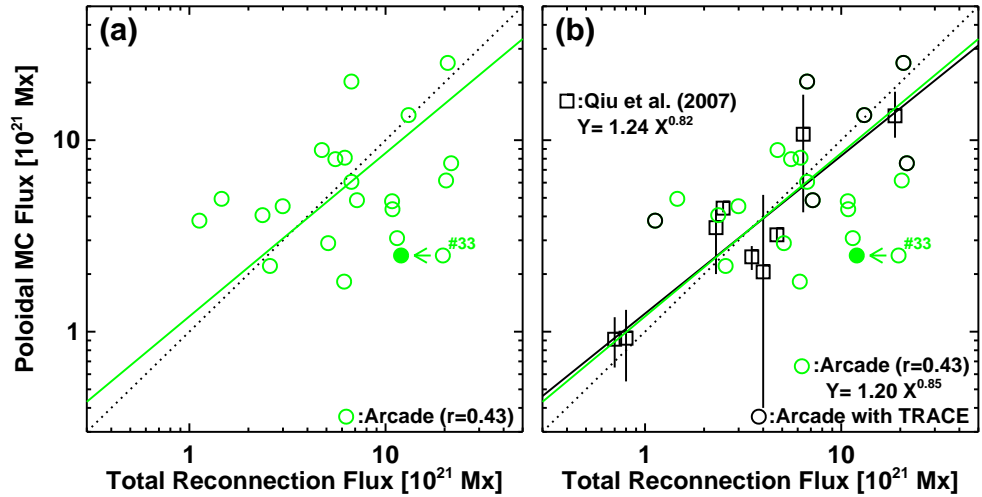

Figure 7. (a) Scatter plot between the RC flux from the arcade method $(\Phi_r)$ and the 1-AU poloidal flux $(\Phi_P)$ in the associated magnetic cloud. The dotted line denotes equal fluxes $(\Phi_P = \Phi_r)$. The correlation coefficient r = 0.43 is larger than the critical value of the Pearson correlation coefficient (0.378 at 95% confidence level). (b) Same as (a) with the data points from Qiu *et al.* (2007) superposed (black squares), showing that the two sets of data yield similar relation between $\Phi_r$ and $\Phi_P$. The vertical black lines indicate the range of values over which the average flux was computed in Qiu *et al.* (2007). We have shown event No. 33 with the uncorrected (open circle) and corrected (filled circle) fluxes. Although this event follows the trend of the scatter plot, we have not included it for the reasons given in the text.

We think Qiu *et al.* (2007) had a misidentification of the solar eruption in this event as described in Figure 6, which shows the PEA and the source region of the eruption from *Yohkoh/Soft X-ray Telescope*, SOHO/EIT, MDI magnetogram, and GOES light curve. This event has large uncertainties both at the Sun and at 1 AU. The CME of interest was a halo (likely to reach Earth) and first appeared in the LASCO FOV at 12:30 UT on 26 April 2001. The CME was associated with a gradual flare with a soft X-ray flare class of M1.5 that resulted in a large PEA. The centroid of the arcade was at N20W05. Roughly after the peak of the M1.5 gradual flare, an impulsive flare (M8.9) started at ~13:04 UT and peaked at 13:13 UT. The M8.9 flare originated about 26° to the west of the centroid of the arcade. The M8.9 flare was not associated with the halo CME. It was associated with a narrow CME that first appeared in the LASCO FOV at 13:31 UT and was heading to the west. Qiu *et al.* (2007) used this flare to get the flare RC flux, but it is unlikely that the 13:31 UT CME would have arrived at Earth. Both the M1.5 and M8.9 eruptions were from the same magnetic complex consisting of AR 9433 surrounded by weaker magnetic flux regions. Figure 6 shows that AR 9433 was located between the arcade centroid and the impulsive flare source. AR 9433 was located under the PEA, but to its western part. The RC flux from the impulsive source was reported by Qiu *et al.* (2007) as $0.7\times10^{21}$ Mx, more than an order of magnitude smaller than that of the M1.5 gradual flare ($12.0\times10^{21}$ Mx after eliminating the flux due to AR 9433).

A similar case of multiple events was observed on 11 April 2004 (No. 49). At the solar source (AR 10696), there were three M-class flares in quick succession (see http://www.lmsal.com/solarsoft/last_events_20041106_1019/index.html for details). The first one was an M9.3 impulsive flare (00:11 to 00:42 UT) from N08E05 with no obvious CME association. The second M5.9 flare (00:44 to 01:10 UT) occurred almost in the same location (N10E06) as the first one, but was associated with a halo CME (Gopalswamy, Yashiro, and Akiyama, 2006, their Figure 4) with a speed of ~818 km/s appearing at 01:31 UT in the LASCO FOV. The third flare was of X-ray class M3.6 (01:40 to 02:08 UT) that occurred to the west and south of the previous flares, at N07E00. This flare was associated with a faster (1111 km/s) partial halo CME, which appeared at 02:06 UT in the LASCO FOV. It quickly caught up with the previous CME. The 1-AU MC has been determined to be due to the second CME (Gopalswamy *et al* 2010). Although the arcade appears to be a single one, it is possible to separate it into individual sections by looking at the time evolution. The arcade corresponding to the second CME had an RC flux of $6.70\times10^{21}$ Mx, while the combined arcade yielded a flux of $14.85\times10^{21}$ Mx. Both Hα and TRACE observations were available for this event, which yielded reasonable values of the RC flux, especially when the two sets of observations were combined (see Figure 4a).

### 3.2 Comparison between the RC flux and the MC Poloidal Flux

We now consider how the RC flux from the arcade method relates to the poloidal flux of the associated MCs listed in Table 1. Figure 7a shows a scatter plot between $\Phi_r$ and $\Phi_P$ (we drop the subscript "A" in $\Phi_{rA}$ for simplicity). We see a modest correlation, with equal number of data points on either side of the equal-fluxes ($\Phi_r = \Phi_P$) line. A linear least-squares fit to $\Phi_r$–$\Phi_P$ pairs (on a logarithmic scale) gives the regression equation,

$$\Phi_P = 1.20(\Phi_r)^{0.85}. \quad\quad\quad\quad\quad\quad\quad\quad\quad\quad\quad\quad\quad\quad\quad\quad\quad\quad\quad\quad\quad\quad\quad\quad (4)$$

The correlation coefficient (r = 0.43) is higher than the critical value for Pearson correlation (0.378 at 95% confidence level). Thus the $\Phi_P$–$\Phi_r$ relationship is very similar to Equation (1), obtained by Qiu *et al.* (2007) for 9 events. The coefficient of $\Phi_r$ in Equation (4) is only 7.1% higher than that in Equation (1). Similarly, the exponent of $\Phi_r$ is only 3.6% larger than that in Qiu *et al.* (2007) result. More importantly, the RC flux used in Equation (1) was from the ribbon method, whereas it is from the arcade method in Equation (4). Thus we confirm the result presented in Figure 4 *viz.*, the RC fluxes obtained from the arcade and ribbon methods lead to similar correlation with the MC poloidal flux. In addition, we see that the flare RC flux at the Sun is closely related to the poloidal flux of the associated MC, especially close to the $\Phi_r = \Phi_P$ line.

The close correspondence between Qiu *et al.* (2007) and our result may be due to a combination of several circumstances. Since Qiu *et al.*, used a flux rope length of 1 AU instead of 2 AU used by us, there may be an overestimate of $\Phi_P$ in our case. If we also overestimated the RC flux in the arcade method, then these two might have balanced to yield the same relationship between $\Phi_P$ and $\Phi_r$. However, Qiu *et al.* (2007) used $\Phi_P$ values that are from three different sources. Sometimes, the values differed by an order of magnitude for a given event. Therefore, the comparison between our $\Phi_P$ and that of Qiu *et al.* (2007) may not be appropriate. When we checked our $\Phi_P$ values with those in Qiu *et al.* (2007) for five events that are also in our list (Nos. 21, 23, 32, 45, and 53), we found that our values are only slightly larger, by a factor 1.3. The reconnection flux from the ribbon method obtained by Qiu *et al.* (2007) in three of the five cases were smaller than our reconnection flux from the ribbon method. Therefore, it is possible that this underestimate and their use of 1 AU for flux rope length balanced out. We also found that one of the data points (10 April 2001) differs between their Table 4 and their Figure 8c. When we used their table values, we get a new line with the same exponent, but with a slightly higher coefficient (1.24 *vs*. 1.12). Because of the small sample available for comparison, it is difficult to derive strong conclusions on the similarity between Equations (1) and (4).

It must be noted that sometimes the association between the solar and interplanetary events may be incorrect. For instance, in the case of the 10 April 2001 event (No. 33) discussed in section 3.1, the small value of RC flux obtained by Qiu *et al.* (2007) matched with the MC poloidal flux at 1 AU ($1.25 \times 10^{21}$ Mx). We think this is fortuitous because the CME-MC association was not correct. Moreover, the MC was quite complex with different authors giving different durations ranging from 11 to 60 hours (see Table 1 of Qiu *et al.*, 2007). In the CDAW list, the MC was reported to have a duration of ~ 11 hours. However, there were ICME material with low proton temperature before and after the MC interval. Our poloidal flux value was $2.5 \times 10^{21}$ Mx (since we use a flux rope length of 2 AU), which is lower by a factor of 4.8 than the RC flux from the arcade. Because of the uncertainties surrounding the 26 April 2001 event, we did not include them in the correlation, although the data points shown in Figure **7** are consistent with the rest of the events.

## 4. Discussion

We presented a new method to estimate the flare RC flux based on post eruption arcades and photospheric magnetic field underlying the arcades. We have shown that (i) half the magnetic flux under post eruption arcades is a reasonable representation of the flare RC flux (Figure 4), and (ii) the RC flux obtained from the arcade method and the poloidal flux of the associated magnetic cloud at 1 AU are correlated significantly (Figure 7). The relationship is very similar to the one obtained by Qiu *et al.* (2007).

One of the major advantages of the new method is that it requires just one image that shows the "mature" arcade in EUV, X-rays, or even microwaves (see *e.g.*, Hanaoka *et al.*, 1994; Gopalswamy *et al.*, 2013). Typically this image corresponds to the decay phase of flares. The availability of observations of post eruption arcades is generally better than that of flare ribbons. We see that only ten of the 23 eruptions from Cycle 23 had usable ribbon observations (from Hα and TRACE 1600 Å). On the other hand, 21 of the 23 eruptions had usable arcade observations. Thus ~3 times more events are usable for the arcade method than for the ribbon method in Cycle 23. The robustness of the arcade method is important for space weather applications.

Hu *et al.* (2014) extended the study of Qiu *et al.* (2007) to include ten more events from Cycle 24, thus doubling their sample. They did find the continued correlation between the flare RC flux and MC poloidal flux, but with a pronounced deviation from the equal-fluxes line. Their sample size (19) is now similar to ours (21), but their solar source locations do not meet our longitude criterion (within ±15º) for all but two events. One of the two events did not have the RC flux computed. The only event matching our criterion (23 May 2010 eruption from N19W12) was reported to have $\Phi_r = 0.3 \times 10^{21}$ Mx and $\Phi_P = 0.83 \times 10^{21}$ Mx (Hu *et al.*, 2014). Substituting this $\Phi_r$ into Equation (4), we get $\Phi_P = 0.43 \times 10^{21}$ Mx, which is within a factor of two from Hu *et al.*'s (2014) $\Phi_P$ value. We note that Hu *et al.* (2014) considered only the Grad-Sha**f**ranov (GS) method of flux rope fitting rather than the force-free (FF) fitting we used. In Qiu *et al.*'s (2007) Table 4, all the poloidal flux values from the FF method were higher than those from the GS method – by a factor of ~1.6 on average. In our case, estimation of the poloidal flux was possible for 13 of the 21 events (Möstl 2014, private communication). In all 13 cases, the FF method gave a poloidal flux higher by a factor of five on average. Similar differences were also found in the five events in Qiu *et al.* (2007) that overlapped with ours: the GS fluxes were smaller by factors of 5.8 than Qiu *et al.* FF values and by a factor of 7.6 than our values. Since we have good information on magnetic clouds of Cycle 24 (Gopalswamy *et al.*, 2015), we will add more data points to the current 21 for better statistics and report the results elsewhere.

We considered only eruptions from the disk center that were observed as MCs at 1 AU. However, there are many disk-center eruptions that are not observed as MCs. Propagation effects such as deflection in the corona (Xie, Gopalswamy and St. Cyr, 2013; Mäkelä *et al.*, 2013) seem to be responsible for the non-cloud appearance of these events at 1 AU, even though there is no difference in the source properties of cloud and non-cloud ICMEs (Gopalswamy *et al.*, 2013;

Yashiro *et al.,* 2013). Therefore, it is possible to estimate the expected poloidal flux of the non-cloud ICMEs at 1 AU from the RC flux. Marubashi *et al. (*2015) were able to fit a flux rope to all but three of the non-cloud ICMEs in the CDAW list. A comparison between the RC flux and the poloidal flux from Marubashi *et al. (*2015) may provide a clue to understand why the Lepping *et al. (*1990) force-free fitting did not recognize flux rope signatures.

**5. Summary and Conclusions**

We investigated a set of 21 solar eruptions originating from within ±15º of the disk center and the associated magnetic clouds from Solar Cycle 23 to demonstrate a new method of measuring the flare RC flux. We measured the RC flux by combining observations of post eruption arcades in EUV and line of sight photospheric magnetic fields. We also measured the RC flux from Hα and TRACE 1600 Å observations using the flare ribbon method. We found that the RC flux obtained from the two methods agreed quite closely. The RC flux from the arcade method is slightly larger than that from the ribbon method. This is mostly due to insufficient flare ribbon data. Occasionally there may be large flux close to the polarity inversion line that does not participate in the reconnection and hence can cause an overestimate of the arcade RC flux.  We also computed the poloidal flux of the associated magnetic clouds at 1 AU and found it to be approximately equal to the RC flux. This result is consistent with the idea that the flux ropes are formed during eruptions and any pre-existing flux rope needs to be rather small in the set of events we considered.

**Acknowledgements**

We thank the ACE, *Wind* and SOHO teams for providing the data on line. SOHO is a project of international collaboration between ESA and NASA. We thank C. Möstl for providing poloidal flux of 13 magnetic clouds using the Grad-Shafranov method.  Work supported by NASA's Living with a Star Program.

**Disclosure of Potential Conflicts of Interest** The authors declare that they have no conflicts of interest.

Table 1. List of solar eruptions that were associated with magnetic clouds at Earth

| No | MC Date UT | CME Date UT | Flare UT | Class | Loc. | $A_a$ | $<B_a>$ | $\Phi_{rA}$ | $A_R$ | $<B_R>$ | $\Phi_{rR}$ | $\Phi_p$ |
|---|---|---|---|---|---|---|---|---|---|---|---|---|
| 02 | 1997/05/15 01:15 | SOL1997-05-12T05:30 | 04:42 | C1.3 | N21W08 | 4.25 | 111.1 | 2.36 | $1.70^H$ | 171.5 | 1.46 | 4.06 |
| 09 | 1999/04/16 11:10 | SOL1999-04-13T03:30 | 01:45 | B4.3 | N16E00 | 10.38 | 28.1 | 1.46 | $2.06^{aH}$ | 46.8 | 0.48 | 4.94 |
| 16 | 2000/02/20 21:00 | SOL2000-02-17T21:30 | 20:17 | M1.3 | S29E07 | 5.57 | 107.4 | 2.99 | $3.24^H$ | 138.6 | 2.25 | 4.52 |
| 19 | 2000/07/15 14:18 | SOL2000-07-14T10:54 | 10:03 | X5.7 | N22W07 | 7.50 | 349.4 | 13.10 | $5.64^{TaH}$ | 446.9 | 12.61 | 13.52 |
| 21 | 2000/07/28 06:39 | SOL2000-07-25T02:43 | 02:43 | M8.0 | N06W08 | 0.93 | 240.9 | 1.13 | $0.82^T$ | 310.3 | 1.27 | 1.90 |
| 23 | 2000/08/11 18:51 | SOL2000-08-09T16:30 | 15:19 | EP | N20E12 | 7.21 | 172.0 | 6.20 | --- | --- | --- | 8.10 |
| 24 | 2000/09/17 17:00 | SOL2000-09-16T05:18 | 04:06 | M5.9 | N14W07 | 3.72 | 298.1 | 5.55 | --- | --- | --- | 7.96 |
| 26 | 2000/10/12 22:36 | SOL2000-10-09T23:50 | 23:19 | C6.7 | N01W14 | 15.94 | 64.0 | 5.10 | $3.08^{aH}$ | 125.1 | 1.93 | 2.90 |
| 27 | 2000/11/06 09:20 | SOL2000-11-03T18:26 | 18:35 | C3.2 | N02W02 | 35.85 | 113.5 | 20.35 | --- | --- | --- | 6.16 |
| 32 | 2001/04/11 16:19 | SOL2001-04-10T05:30 | 05:06 | X2.3 | S23W09 | 7.34 | 194.9 | 7.15 | $5.75^{H+T}$ | 299.9 | 8.62 | 4.86 |
| 33 | 2001/04/28 05:02 | SOL2001-04-26T12:30 | 11:26 | M1.5 | N20W05 | 16.45 | 145.9 | 12.00 | $3.86^{aH}$ | 167.2 | 6.45 | 2.50 |
| 36 | 2002/03/18 13:13 | SOL2002-03-15T23:06 | 22:09 | M2.2 | S08W03 | 12.46 | 183.8 | 11.45 | $1.18^{aH}$ | 363.1 | 4.28 | 3.08 |
| 37 | 2002/04/17 11:01 | SOL2002-04-15T03:50 | 03:05 | M1.2 | S15W01 | 4.78 | 452.1 | 10.80 | $0.81^H$ | 1137.1 | 4.58 | 4.80 |
| 39 | 2002/05/18 19:51 | SOL2002-05-16T00:50 | 00:11 | C4.5 | S23E15 | 8.36 | 113.2 | 4.74 | ---- | ---- | --- | 8.88 |
| 43 | 2002/08/01 05:10 | SOL2002-07-29T12:07 | 10:27 | M4.7 | S10W10 | 2.40 | 511.5 | 6.15 | $0.64^{aT}$ | 426.4 | 1.35 | 1.83 |
| 44 | 2003/08/17 13:40 | SOL2003-08-14T20:06 | 17:12 | wave | S10E02 | 4.16 | 522.1 | 10.85 | ---- | ---- | --- | 4.36 |
| 45 | 2003/10/29 06:00 | SOL2003-10-28T11:30 | 11:00 | X17.2 | S16E08 | 11.94 | 347.5 | 20.75 | $7.05^T$ | 592.1 | 20.88 | 25.32 |
| 46 | 2003/10/30 16:20 | SOL2003-10-29T20:54 | 20:37 | X10.0 | S15W02 | 8.71 | 496.1 | 21.60 | $3.96^{TaH}$ | 698.4 | 13.82 | 7.58 |
| 49 | 2004/11/09 09:05 | SOL2004-11-06T02:06 | 01:40 | M3.6 | N09E05 | 4.12 | 325.5 | 6.70 | $0.98^{H+T}$ | 1009.8 | 4.93 | 6.06 |
| 53 | 2005/05/15 02:19 | SOL2005-05-13T17:12 | 16:13 | M8.0 | N12E11 | 4.88 | 274.8 | 6.70 | $4.06^{H+T}$ | 353.6 | 7.19 | 20.24 |

| 54 | 2005/05/20 03:34 | SOL2005-05-17T03:26 | 02:31 | M1.8 | S15W00 | 2.45 | 210.6 | 2.58 | ---- | ---- | --- | 2.22 |

Notes.

Flare date is the same as the CME date

Under flare class: EP = eruptive prominence, wave = EUV wave

$A_a$ – Area under the post eruption arcade in $10^{19}$ cm$^2$

$<B_a>$ – average photospheric magnetic field strength (G) under the arcade

$\Phi_{rA}$ – RC flux (in $10^{21}$ Mx) obtained from the arcade method (half of the flux passing through area $A_a$)

$A_R$ – Cumulative ribbon area (in $10^{19}$ cm$^2$) from Hα or TRACE 1600 Å observations

$<B_R>$ – average photospheric magnetic field strength (G) under the cumulative ribbon area

$\Phi_{rR}$ – RC flux (in $10^{21}$ Mx) obtained from the ribbon method

$\Phi_p$ – Poloidal flux (in $10^{21}$ Mx) of the magnetic cloud at 1 AU in $10^{19}$ cm$^2$

Superscript H – complete Hα observations

Superscript T – complete TRACE observations

Superscript aH – incomplete Hα observations

Superscript aT – incomplete TRACE observations

Superscript H+T – Complete Hα and TRACE ribbon observations; combined Hα and TRACE information listed

Superscript TaH – Complete TRACE and incomplete Hα ribbon observation; TRACE information listed

Table 2. List of events with RC flux computed from Hα or TRACE 1600 Å ribbons

| ID | CME Date UT | $\Phi_{rA}$ | FrH | $A_{RH}$ | $<B_{RH}>$ | $\Phi_{rRH}$ | FrT | $A_{RT}$ | $<B_{RT}>$ | $\Phi_{rRT}$ | Hα |
|---|---|---|---|---|---|---|---|---|---|---|---|
| 02 | SOL1997-05-12T05:30 | 2.36 | 6 | 1.70 | 171.5 | 1.46 | -- | ---- | ---- | ---- | HIDA |
| 09 | SOL1999-04-13T03:30 | 1.46 | 1 | 2.06 | 46.8 | 0.48[b] | -- | ---- | ---- | ---- | NAOJ |
| 16 | SOL2000-02-17T21:30 | 2.99 | 4 | 3.24 | 138.6 | 2.25 | -- | ---- | ---- | ---- | MLSO |
| 19 | SOL2000-07-14T10:54 | 13.10 | 1 | 1.69[a] | 557.5 | 9.44[b] | 71 | 5.64 | 446.9 | 12.61 | MEUDON |
| 21 | SOL2000-07-25T02:43 | 1.13 | -- | ---- | ---- | ---- | 29 | 0.82 | 310.3 | 1.27 | ----- |
| 26 | SOL2000-10-09T23:50 | 5.10 | 1 | 3.08 | 125.1 | 1.93[b] | -- | ---- | ---- | ---- | NAOJ |
| 32 | SOL2001-04-10T05:30 | 7.15 | 14 | 4.93 | 317.5 | 7.82 | 200 | 4.67 | 306.9 | 7.17 | HIDA |
| 33 | SOL2001-04-26T12:30 | 12.00 | 9 | 3.86[a] | 167.2 | 6.45[b] | -- | ---- | ---- | ---- | OAFA |
| 36 | SOL2002-03-15T23:06 | 11.45 | 1 | 1.18[a] | 363.1 | 4.28[b] | -- | ---- | ---- | ---- | YNAO |
| 37 | SOL2002-04-15T03:50 | 10.80 | 11 | 0.81 | 1137.1 | 4.58 | -- | ---- | ---- | ---- | HIDA |
| 43 | SOL2002-07-29T12:07 | 6.15 | -- | ---- | ---- | --- | 8 | 0.64 | 426.4 | 1.35[b] | ----- |
| 45 | SOL2003-10-28T11:30 | 20.75 | -- | ---- | ---- | --- | 62 | 7.05 | 592.1 | 20.88 | ----- |
| 46 | SOL2003-10-29T20:54 | 21.60 | 3 | 2.21 | 540.9 | 5.99[b] | 101 | 3.96 | 698.4 | 13.82 | MLSO |
| 49 | SOL2004-11-06T02:06 | 6.70 | 7 | 0.92 | 893.5 | 4.11 | 345 | 0.43 | 1168.8 | 2.52 | HIDA |
| 53 | SOL2005-05-13T17:12 | 6.70 | 4 | 3.21 | 372.7 | 5.98 | 454 | 2.89 | 371.5 | 5.36 | BBSO |

Notes.

$\Phi_{rA}$ – RC flux (in $10^{21}$ Mx) obtained from the arcade method (half of the flux passing through area $A_a$)

FrH – Number of Hα frames available for measuring the RC flux

$A_{RH}$ – Cumulative ribbon area (in $10^{19}$ cm$^2$) from Hα observations. Superscript a denotes that only one of the ribbons in the pair was considered for flux measurement

$<B_{RH}>$ – average magnetic field strength (G) within the cumulative Hα ribbon area

$\Phi_{rRH}$ – RC flux (in $10^{21}$ Mx) in units of obtained from Hα ribbons. Suffix b stands for incomplete events. Events without data during flare impulsive phase are considered incomplete

FrT – Number of TRACE 1600 Å frames available for measuring the RC flux

$A_{RT}$ – Cumulative ribbon area from TRACE 1600 Å observations in $10^{19}$ cm$^2$

$<B_{RT}>$ – average magnetic field strength (G) within the cumulative TRACE 1600 Å ribbon area

$\Phi_{rRT}$ – RC flux (in $10^{21}$ Mx) in units of obtained from TRACE ribbons. Superscript b denotes that the RC flux is an underestimate due to lack of sufficient observations

BBSO – Big Bear Solar Observatory (http://www.bbso.njit.edu/Research/FDHA/)

HIDA – Hida Observatory, Kyoto University (http://www.kwasan.kyoto-u.ac.jp/observation/data/index_en.html)

MEUDON – Observatory de Paris, Meudon (http://bass2000.obspm.fr/home.php)

MLSO – Mauna Loa Solar Observatory (http://www2.hao.ucar.edu/mlso/mlso-home-page)

NAOJ – Solar Observatory, NAOJ (http://solarwww.mtk.nao.ac.jp/en/database.html)

OAFA – Observatorio Astronomico Felix Aguilar (http://www.oafa.fcefn.unsj-cuim.edu.ar/hasta/Web/hastasearch.html)

YNAO – Yunnan Astronomical Observatory (http://swrl.njit.edu/ghn_web/latestimg/latestimg.php)